\documentclass[aps,prd,twocolumn,showpacs,showkeys,superscriptaddress, preprintnumbers]{revtex4} 

\usepackage{graphicx}
\usepackage{multirow}
\usepackage{amsmath}
\usepackage{float}
\usepackage{color}
\usepackage{epstopdf}
\usepackage{epsfig}

\font\FermiSmallfont=cmssq8 scaled 1200

\def\LANLppthead#1{
\null 
\begin{center}\vskip -1.0truein{\hbox to 7.5truein {
\hfill
\vbox to 1in {\vfill \FermiSmallfont
              \hbox{#1}
              \vfill}
}}\vskip-0.0truein\end{center}}

\begin{document}

\title{Matter-neutrino resonance in a multi-angle neutrino bulb model}
\preprint{}

\author{Alexey Vlasenko}
\affiliation{Department of Physics, North Carolina State University, Raleigh, NC 27695-8202, USA}
\author{G. C. McLaughlin}
\affiliation{Department of Physics, North Carolina State University, Raleigh, NC 27695-8202, USA}

\begin{abstract}
Simulations of neutrino flavor evolution in compact merger environments have shown that  neutrino flavor, and hence nucleosynthesis, can be strongly affected by the presence of matter-neutrino resonances (MNRs), where there is a cancelation between the matter and the neutrino potential.  Simulations performed thus far follow flavor evolution along a single neutrino trajectory, but self-consistency requires all trajectories to be treated simultaneously, and it has not been known whether MNR phenomena would still occur in multi-angle models.  In this paper, we present the first fully multi-angle calculations of MNR.  We find that familiar MNR phenomena, where neutrinos transform to a greater extent than anti-neutrinos and a feedback mechanism maintains the cancellation between the matter and neutrino potential, still occurs for a subset of angular bins, although the flavor transformation is not as efficient as in the single-angle case.  In addition, we find other types of flavor transformation that are not seen in single-angle simulations.  These flavor transformation phenomena appear to be robust and are present for a wide range of model parameters, as long as an MNR is present.  Although computational constraints currently limit us to models with spherical symmetry, our results suggest that the presence of an MNR generally leads to large-scale neutrino flavor evolution in multi-angle systems.
\end{abstract}


\maketitle

\section{Introduction}

Neutron star - neutron star mergers  produce neutrinos in vast numbers.  In fact a substantial fraction of the gravitational binding energy of two neutron stars can be lost in neutrinos.  The neutrinos produced in such a compact object merger play a role in the dynamics of the object, e.g. \cite{Foucart:2014nda,Perego:2014fma, Deaton:2013sla,Palenzuela:2015dqa,Kiuchi:2015qua}, in the nucleosynthesis in produces, e.g. \cite{Roberts:2010wh,Wanajo:2014wha,Surman:2005kf,Surman:2008qf,Caballero:2011dw,Perego:2014fma,Bauswein:2014vfa,Just:2014fka,Goriely:2015fqa,Foucart:2015gaa,Martin:2015hxa,Foucart:2016rxm,Lippuner:2017bfm},and  in the prospects for forming a jet, e.g. \cite{Janka:1999qu,Nagakura:2014hza,Richers:2015lma,Fujibayashi:2017xsz,Perego:2017fho}.  If a merger were close enough, neutrinos would be detected in the same ways core collapse supernova neutrinos will be detected \cite{Caballero:2009ww,Caballero:2015cpa}.

Many of the neutrino interaction processes relevant to compact object mergers have flavor dependent rates.  For example,  processes that convert protons to neutrons and vice versa proceed in large part through absorption and emission of electron flavor neutrinos.  The ratio of neutrons to protons has a strong effect on element synthesis and flavor transformation has been shown to alter that ratio in winds from compact object mergers \cite{Malkus:2012ts,Malkus:2014iqa,Zhu:2016mwa,Frensel:2016fge,Tian:2017xbr,Wu:2017drk}.  Similarly, the reactions by which merger neutrinos would be detected on earth are the same as those by which supernova neutrinos will be detected and these reactions have significant flavor dependence, e.g. \cite{Scholberg:2012id,Caballero:2015cpa}. Thus, there is strong motivation for understanding well the flavor content of neutrinos produced by compact object mergers and how this flavor content evolves.

Neutrinos can flavor transform both when they are trapped, i.e. changing their momentum through scattering often in the interior of the merger remnant, and when they are free streaming, i.e. scattering rarely above the remnant.  Significant strides have been made into the theory of neutrino flavor transformation in trapped and semi-trapped regime \cite{Strack:2005ux,Vlasenko:2013fja,Volpe:2013uxl}, but as this problem is numerically challenging,  only initial attempts have been made to apply this theory to flavor evolution in astrophysical systems \cite{Cherry:2012zw,Vlasenko:2014bva}.  Also numerically challenging has been the study of the growth of small perturbations located near the beginning of the free streaming region.  The simplified models that have been explored to date suggest that significant growth of these instabilities may occur and further exploration is necesssary \cite{Abbar:2015mca,Abbar:2015fwa,Dasgupta:2016dbv,Wu:2017qpc,Capozzi:2017gqd,Wu:2017qpc}, and new techniques for determining flavor transformation are under study \cite{Armstrong:2016mnt}.

In this work we focus on the free streaming regime in the absence of perturbations.  Neutrino flavor evolution in the free streaming region is governed by a Schr\"odinger-like equation with the Hamiltonian of the form
\begin{eqnarray}
\nonumber
H = H_{\rm VAC}+H_{\rm M}+H_{\nu\nu}
\end{eqnarray}
where $H_{\rm VAC}$ is the vacuum Hamiltonian (the neutrino mass term), $H_{\rm M}$ is the matter potential stemming from coherent forward scattering of neutrinos on the matter background, and $H_{\nu\nu}$ is the neutrino-neutrino interaction potential generated by coherent forward scattering of neutrinos on different trajectories with each other.  
The flavor evolution Hamiltonian is an $N_F\times N_F$ matrix, where $N_F$ is the number of neutrino flavors, and the Hamiltonian for anti-neutrinos differs from that of neutrinos by the sign of the vacuum term.
The matter potential is diagonal in the flavor basis, so in the presence of large matter densities, flavor evolution can be suppressed.  However, even in the presence of large matter density, the nonlinearity of the last term  produces a variety of interesting flavor evolution behaviors under certain conditions.  For example, collective neutrino oscillations may occur in supernovae e.g.  \cite{Duan:2006an,Hannestad:2006nj,Balantekin:2006tg,Duan:2007mv,EstebanPretel:2008ni,Dasgupta:2008cd,Gava:2009pj,Cherry:2011fn} as well as  in the accretion disks and hypermassive neutron stars that can arise from compact object mergers \cite{Dasgupta:2008cu,Malkus:2012ts,Frensel:2016fge,Tian:2017xbr}.  In the presence of non-standard interactions, other types of flavor transformation can occur both in supernovae, e.g. \cite{Stapleford:2016jgz}, and in compact object mergers \cite{Chatelain:2017yxx}, and there may also be the possibility of neutrino spin transformation \cite{Chatelain:2016xva}.

In addition, compact object mergers offer an alternative path for neutrino flavor transformation in the presence of a high matter density.  In certain epochs, the electron anti-neutrino flux exceeds the electron neutrino flux, which leads to a largely flavor-diagonal neutrino-neutrino interaction potential with the opposite sign to the matter potential.  When a cancelation between the neutrino-neutrino and the matter potentials occurs, large-scale flavor evolution can take place.  This is known as a matter-neutrino resonance (MNR)  \cite{Malkus:2014iqa,Vaananen:2015hfa,Wu:2015fga}.

Previous work \cite{Malkus:2012ts,Malkus:2014iqa,Zhu:2016mwa,Frensel:2016fge,Tian:2017xbr} has shown that the presence of MNRs may drastically alter neutrino flavor content in regions of compact object mergers where wind nucleosynthesis takes place.  These studies have followed neutrino flavor evolution along a single trajectory, making the assumption that neutrinos on all other trajectories follow the same flavor evolution, in what is known as the single-angle or single-trajectory approximation.  However, in a real compact object merger, neutrinos on different trajectories pass through different environments and have different histories of flavor evolution, and all trajectories contribute to the neutrino-neutrino potential.  Therefore, a self-consistent calculation must solve for flavor evolution on all trajectories simultaneously.  One calculation of MNR transformation with multiple neutrino trajectories has been performed in \cite{2017arXiv170707692S}, which examined flavor evolution of a neutrino beam with a nonzero opening angle passing through a constant matter profile.  However, flavor evolution due to MNR in fully multi-angle compact object environments has largely remained an open problem.

Neutron star - neutron star mergers are now thought to form hypermassive neutron stars surrrounded by accretion disks, with a considerable fraction of the neutrino emission originating from the hypermassive neutron star. In a full multi-angle flavor evolution calculation, neutrinos may be labeled by position, time, propagation angle and neutrino energy.  Unfortunately, the high dimensionality, together with high resolution required in some of the dimensions,  makes the full flavor evolution problem computationally unfeasible for the forseeable future.  Therefore, we take an inital step by 
considering a spherically symmetric model as an approximation for the hypermassive neutron star.  Even though real compact object mergers are not spherically symmetric because they also have a surrounding accretion disk, this model shares some key features with the full problem, and allows us for the first time to examine whether MNRs may have an effect on flavor evolution in a fully self-consistent calculation that includes multiple neutrino trajectories.

In this paper, we explore multi-angle effects on the MNR phenomnenon in astrophysically motivated scenarios.  In Sec. \ref{sec:description} we describe the model and set up the calculations as well as explain our numerical implementation. In Sec. \ref{sec:results} we discuss our results for several different scenarios and in Sec. \ref{sec:conclusions} we present our conclusions.

\section{Description of the Model}
\label{sec:description}
\begin{figure}
\includegraphics[width=2.8in]{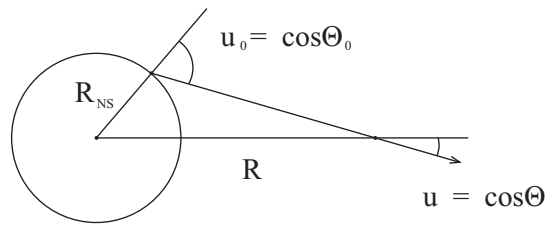}
\caption{Geometry of the neutrino bulb model. The distance from the center to the neutrino emission surface is represented by $R_{NS}$, the distance from the center to the current propagation distance is given by $R$.  The angle $\Theta_{0}$ represents the emission angle of the neutrino: a neutrino emitted perpendicular to the emission surface has $\Theta_0 = 0$, while a neutrino emitted trangentially to the surface has $\Theta_0= \pi/2$.  The propagation angle is represented by $\Theta$.} 
\label{geometry}
\end{figure}

To describe the evolution of the neutrinos we use a model which is almost identical to the usual treatment of supernova flavor evolution developed in, {\it e.g.}  \cite{Duan:2006an}, sometimes called the bulb model.  However, to capture the essential physical difference that creates MNR conditions, we make the crucial change that the initial neutrino spectrum contains an excess of electron anti-neutrinos.  In terms of geometry, this model, which is illustrated schematically in Fig. \ref{geometry}, is spherically symmetric, with neutrinos that are emitted semi-isotropically from a spherical surface of fixed radius (the neutrinosphere) and propagate outward.  The outward arrow in Fig \ref{geometry} shows just one neutrino trajectory, labeled by the cosine of the emission angle $u_0$, but in our simulations, all outward-directed neutrino trajectories, $0\leq u_0\leq 1$, are tracked simultaneously: this is  multi-angle aspect of the calculation.  We follow the standard treatment where the flavor field is assumed to be static in time and backscattered neutrinos are neglected, so the equations for neutrino flavor evolution can be written as an initial value problem in radius.  At a given radius, the flavor state is function of energy and emission angle.  The energy and emission angle are binned, giving a large set of nonlinearly coupled differential equations that simultaneously describe the flavor evolution of the entire system. We give a more detailed description of our calculations in the remainder of this section.  

\subsection{The Hamiltonian in the bulb model}

We begin by describing these nonlinearly coupled differential equations and explaining a key point of this paper, which is that the location of the MNR, approximately defined as the point where the neutrino potential and the matter potential are equal and opposite, will differ for different neutrinos depending on their emission angle.

The flavor state for a given energy and angle can be represented with a pair of $N_F\times N_F$ flavor density matrices, one for neutrinos and one for anti-neutrinos.  For simplicity, we consider only two flavors, labeled as $\nu_e$ and $\nu_x$, although the extension to three or more flavors is straightforward.  For two flavors, the density matrices take on the form:
\begin{eqnarray}
\rho = \left(\begin{array}{cc}\rho_{ee} & \rho_{ex} \\ \rho_{xe} & \rho_{xx}\end{array}\right)\ \ \ \ \ \ \ \bar{\rho}=\left(\begin{array}{cc}\bar{\rho}_{ee} & \bar{\rho}_{ex} \\ \bar{\rho}_{xe} & \bar{\rho}_{xx}\end{array}\right)
\end{eqnarray}
The density matrices are Hermitean, so that $\rho_{xe}=\rho_{ex}^*$, and $\rho_{ee}$, $\rho_{xx}$ are real.  Here, they are defined so that the real diagonal elements correspond to the occupation numbers for the electron flavor and the $\mu$/$\tau$ flavor, for a particular energy and angular bin.

The matrices $\rho$ and $\bar{\rho}$ follow Schr\"odinger-like evolution equations along each trajectory:
\begin{eqnarray}
\frac{d\rho\left(E,u_0\right)}{ds}=-i\left[H^+\left(E,u_0\right),\rho\left(E,u_0\right)\right]\nonumber\\
\frac{d\bar{\rho}\left(E,u_0\right)}{ds}=-i\left[H^-\left(E,u_0\right),\bar{\rho}\left(E,u_0\right)\right]
\label{eq2}
\end{eqnarray}
where the trajectories are labeled by the neutrino or anti-neutrino energy $E$, and the cosine of the emission angle $u_0=\cos\theta_0$, which corresponds to the direction of neutrino propagation at the neutrinosphere as shown in Fig.~\ref{geometry}.  The variable $s$ is the neutrino propagation distance.   The Hamiltonian for neutrinos and anti-neutrinos consists of a vacuum part, a matter potential, and a neutrino-neutrino interaction potential:
\begin{eqnarray}
H^\pm = \mp H_{\rm VAC}+H_{\rm M}+H_{\nu\nu}
\label{eq3}
\end{eqnarray}
Due to the presence of the commutators on the right-hand side, only the traceless part of the Hamiltonian affects the flavor evolution, and the traces of the density matrices are conserved.  For two flavors, we can decompose the density matrices as
\begin{eqnarray}
\rho = \rho_T{\rm\bf 1}+\rho_i\sigma^i = \rho_T{\rm \bf 1}+\vec{\rho}\cdot\vec{\sigma}
\end{eqnarray}
and similarly for $\bar{\rho}$ and $H$.  Here, $\rho_T$ is proportional to the total number of neutrinos of all species on a given trajectory, while the components of $\vec{\rho}$ describe the flavor asymmetry.

In this notation, 
\begin{eqnarray}
\rho_T&=& \frac{1}{2}\left(\rho_{ee}+\rho_{xx}\right)
\ \ 
\rho_3 = \frac{1}{2}\left(\rho_{ee}-\rho_{xx}\right)
\nonumber\\
\rho_1 &=& {\rm Re}\left(\rho_{ex}\right)
\ \ \ \ \ \ \ \ \ 
\rho_2 = {\rm Im}\left(\rho_{ex}\right)
\end{eqnarray}

For two flavors, the traceless part of the vacuum Hamiltonian can be written in the flavor basis as:
\begin{eqnarray}
H_{\rm VAC} = \frac{\Delta m^2}{4E}\left(\begin{array}{cc}\cos 2\theta_{\rm VAC} & \sin 2\theta_{\rm VAC} \\ \sin 2\theta_{\rm VAC} & -\cos 2\theta_{\rm VAC}\end{array}\right)
\end{eqnarray}
where $\theta_{\rm VAC}$ is the vacuum mixing angle.  In components,
\begin{eqnarray}
H_{{\rm VAC},1} &=& \frac{\Delta m^2}{4E}\sin 2\theta_{\rm VAC}
\nonumber\\
H_{{\rm VAC},2} &=& 0
\nonumber\\
 H_{{\rm VAC},3} &=& \frac{\Delta m^2}{4E}\cos 2\theta_{\rm VAC}
\end{eqnarray}
With the sign convention as given by Eqn.~\ref{eq3}, a positive sign for $\cos 2\theta_{\rm VAC}$ corresponds to the normal mass hierarchy, and a negative sign to the inverted.  In this paper, we use the normal hierarchy.

 The traceless part of the matter potential is
\begin{eqnarray}
H_{\rm M} = \sqrt{2}G_F n_B Y_e\left(\begin{array}{cc}1 & 0 \\ 0 & -1\end{array}\right)
\end{eqnarray}
where $n_B$ is the baryon number density and $Y_e$ is the electron to baryon ratio.  While $n_B$ and $Y_e$ profiles may be obtained from hydrodynamic simulations of compact object mergers,
we note that the matter density typically decreases roughly as $1/R^3$, and adopt the following
for the matter potential profile:
\begin{eqnarray}
H_{\rm M} = \sqrt{2}G_F n_{\rm B,NS} Y_e\frac{R_{\rm NS}^3}{R^3}\left(\begin{array}{cc}1 & 0 \\ 0 & -1\end{array}\right)
\end{eqnarray}
where $R_{\rm NS}$ is the neutrinosphere radius, $Y_e$ is the electron fraction and $n_{\rm B,NS}$ is the baryon number density at the neutrinosphere.  In components, 
\begin{eqnarray}
H_{M,3} &=& \sqrt{2}G_F n_{\rm B,NS} Y_e\frac{R^3_{\rm NS}}{R^3}
\nonumber\\
H_{M,1}&=&H_{M,2}=0
\label{eq:Hm}
\end{eqnarray}

We follow \cite{Duan:2006an} in writing the neutrino potential as
\begin{eqnarray}
H_{\nu\nu} = H_0-u H_R
\label{eq:Hnunu}
\end{eqnarray}
where $u$ is the cosine of the propagation angle of the neutrino of interest at a particular radius, as shown in Fig.~\ref{geometry}.  The components of the neutrino Hamiltonian $H_0$ and $H_R$ are proportional to the neutrino lepton number density and radial lepton number flux and can be written as 
\begin{eqnarray}
H_0 &=& \sqrt{2}G_F\int_0^\infty\frac{E^2dE}{4\pi^2}\int_{u_{\rm MIN}}^1 du^{\prime}\left[\rho-\bar{\rho}\right]
\nonumber\\
H_R &=& \sqrt{2}G_F\int_0^\infty\frac{E^2dE}{4\pi^2}\int_{u_{\rm MIN}}^1 u^{\prime}\ du^{\prime} \left[\rho-\bar{\rho}\right]
\label{eq:H0HR}
\end{eqnarray}
The quantity $u^{\prime}$ indicates the cosine of the angle of the background neutrino.
For a trajectory with a given emission angle $u_0=\cos\theta_0$, the cosine of the propagation angle at radius $R$ is
\begin{eqnarray}
u\left(u_0,R\right) = \sqrt{1-\frac{R_{NS}^2}{R^2}\left(1-u_0^2\right)}
\label{eq8b}
\end{eqnarray}
Thus if a neutrino is emitted perpendicularly to the neutrino sphere, its propagation angle will remain the same as its emission angle, $u = u_0 = 1$.  At any given radius, the largest angle, $u_{\rm MIN}$, with which neutrinos arrive at a distance $R$ above the neutrino emission surface occurs for $u_0 = 0$.
Eqns. \ref{eq:H0HR} now be written as
\begin{eqnarray}
H_0
&=& \sqrt{2}G_F\frac{R_{NS}^2}{R^2}\int_0^\infty\frac{E^2dE}{4\pi^2}\int_0^1 \frac{u_0^{\prime}}{u^{\prime}}du^{\prime}_0\left[\rho-\bar{\rho}\right]
\nonumber\\
H_R &=&\sqrt{2}G_F\frac{R_{NS}^2}{R^2}\int_0^\infty\frac{E^2dE}{4\pi^2}\int_0^1 u_0^{\prime}du^{\prime}_0\left[\rho-\bar{\rho}\right]
\label{eq9}
\end{eqnarray}
where the intergals are now performed over the emission angle $u_0^{\prime}$ of the background neutrinos.

The neutrino-neutrino interaction potential, $H_{\nu\nu}$ at large distance $R >> R_{NS}$, decreases as $1/R^4$.  This comes from a combination of the  geometric flux dilution, seen as the $1/R^2$ term outside both integrals and the 
the structure of the neutrino forward scattering potential which causes a cancellation of the first order terms in Eq. \ref{eq:Hnunu}.  On the other hand, $H_{\rm M}$ decreases as $1/R^3$.  Therefore, if we begin with initial conditions where the neutrino potential is of opposite sign and larger magnitude than the matter potential, at some larger radius the neutrino and matter potentials will cancel, and an MNR will occur.  

As mentioned earlier, a key point of this paper is that the location of the MNR will differ for different neutrinos depending on the emission angle. This can be understood as follows.  Neutrinos emitted vertically, with $\theta_0 = 0$ and $u_0 = 1$, have the smallest neutrino potential and will therefore cross the MNR first, while neutrinos emitted tangentially, with $\theta_0 = \pi/2$ and $u_0 = 0$, have the largest neutrino potential and will encounter the MNR farther out, where the neutrino potential has decreased further compared to the matter potential.  This point is true of any multi-angle model, with or without spherical symmetry, and contrasts the single-angle approximation, where all neutrinos cross the resonance at the same time.

\subsection{Equations of motion for the single-energy and small-angle approximations}

Eqn.s.~\ref{eq2} conserve the trace of the density matrices $\rho_0, \bar{\rho}_0$ as well as the magnitude of the traceless part, $\left|\rho\right|=\sqrt{\rho_1^2+\rho_2^2+\rho_3^2}$.  We can therefore write $\rho = \rho_T{\bf \rm 1}+\left|\rho\right|\hat{\rho}$, and similarly for $\bar{\rho}$.  In this notation, the EOMs for the density matrix of each angle and energy become
\begin{eqnarray}
\rho_T, \bar{\rho}_T &=& {\rm constant}\ \ \ \ 
\left|\rho\right|, \left|\bar{\rho}\right| ={\rm constant}
\label{eq:conserved}
\end{eqnarray}
\begin{eqnarray}
\frac{d\hat{\rho}}{ds}&=&\vec{H}^+\times\hat{\rho}
\ \ \ \ \ \ \ \ \ \ 
\frac{d\hat{\bar{\rho}}}{ds}=\vec{H}^-\times\hat{\bar{\rho}}
\label{eq11}
\end{eqnarray}

The initial conditions for the flavor unit vectors, assuming that there is an excess of both electron neutrinos and anti-neutrinos over the $\mu/\tau$ flavor and that the neutrinos start in a flavor eigenstate, are
\begin{eqnarray}
\hat{\rho}\left(R_0\right) = \hat{\bar{\rho}}\left(R_0\right) = \hat{e}_3
\end{eqnarray}
where $R_0$ is the radius at which the calculation begins.  Throughout this paper, $R_0$ is chosen to be above the neutrinosphere, but sufficiently below the region where flavor transformation takes place that changing the value has no impact on the results of the calculation.

In our calculations, we use the single energy approximation, which reduces the computational complexity of the problem and improves numerical stability, while retaining many of the important features.  

In our  single-energy model, we assume that the flavor evolution for neutrinos and anti-neutrinos of all energies is the same, and integrate all density matrices over energy.  Thus, the energy-integrated analogues of the conserved quantities of Eqn.s.~\ref{eq:conserved} are
\begin{eqnarray}
\rho_T = \int\frac{E^2 dE}{2\pi^2}\frac{1}{2}\left(\rho_{ee}+\rho_{xx}\right)
\nonumber\\
\bar{\rho}_T = \int\frac{E^2 dE}{2\pi^2}\frac{1}{2}\left(\bar{\rho}_{ee}+\bar{\rho}_{xx}\right)
\nonumber\\
\left|\rho\right| = \left|\int\frac{E^2 dE}{2\pi^2}\frac{1}{2}\left(\rho_{ee}-\rho_{xx}\right)\right|
\nonumber\\
\left|\bar{\rho}\right| = \left|\int\frac{E^2 dE}{2\pi^2}\frac{1}{2}\left(\bar{\rho}_{ee}-\bar{\rho}_{xx}\right)\right|
\end{eqnarray}

We further define two quantites as follows:
\begin{eqnarray}
k &=& \sqrt{2}G_F\int\frac{E^2 dE}{4\pi^2}\frac{1}{2}\left(\rho_{ee,0}-\rho_{xx,0}\right)
\nonumber\\
\alpha &=& \frac{\int E^2 dE\left(\bar{\rho}_{ee,0}-\bar{\rho}_{xx,0}\right)}{\int E^2 dE\left(\rho_{ee,0}-\rho_{xx,0}\right)}.
\end{eqnarray}
The quantity $k$ gives the overall scale of the neutrino potential, while the quantity $\alpha$ gives the anti-neutrino contribution to the potential relative to the neutrino contribution.  MNR can occur when $\alpha > 1$.

The components of the neutrino Hamiltonian corresponding to Eqn.~\ref{eq9} can now be  written as
\begin{eqnarray}
\vec{H}_0 = k\frac{R_{\rm NS}^2}{R^2}\int_0^1 \frac{u_0^{\prime}}{u^{\prime}} du_0^{\prime}\left[\hat{\rho}-\alpha\hat{\bar{\rho}}\right]
\nonumber\\
\vec{H}_R = k\frac{R_{\rm NS}^2}{R^2}\int_0^1 u_0^{\prime} du_0^{\prime}\left[\hat{\rho}-\alpha\hat{\bar{\rho}}\right]
\end{eqnarray}
The matter potential is unchanged in the single energy approximation.  The energy in the vacuum Hamiltonian is replaced with $ \langle E \rangle $ or $\langle \bar{E}\rangle$, the average neutrino or anti-neutrino energy.

In addition to adopting the single-energy approximation, we will limit ourselves to situations where MNR takes place at $R >> R_{\rm NS}$ and adopt the small-angle approximation.  This is advantageous because at large values of $R$, the two terms in Eqn.~\ref{eq:H0HR} are nearly equal.  Since one is subtracted from the other, this can lead to numerical instability.  Using Eqn.s.~\ref{eq:H0HR}-\ref{eq9}, expanding $u$ to leading order in in $R_{\rm NS}/R$  and adopting the single-energy approximation, we obtain
\begin{eqnarray}
\vec{H}_{\nu \nu} \approx k\frac{R_{\rm NS}^4}{R^4}\left[\left(1-\frac{1}{2}u_0^2\right)\vec{\Phi}_1-\frac{1}{2}\vec{\Phi}_3\right]
\label{eq:hnunu}
\end{eqnarray}
where
\begin{eqnarray}
\vec{\Phi}_N \equiv \int_0^1 du_0' u_0'^N\left(\hat{\rho}-\alpha\hat{\bar{\rho}}\right)
\end{eqnarray}
It is convenient to represent the emission angle by $v\equiv u_0^2$ instead of $u_0$.  With this change of variables,
\begin{eqnarray}
\vec{\Phi}_1=\frac{1}{2}\int_0^1 dv'\left(\hat{\rho}-\alpha\hat{\bar{\rho}}\right)
\nonumber\\
\vec{\Phi}_3=\frac{1}{2}\int_0^1 v'dv'\left(\hat{\rho}-\alpha\hat{\bar{\rho}}\right)
\end{eqnarray}
In the small-angle approximation, neutrinos travel nearly radially outward.  This, together with the fact that the angular bin label $u_0$ is conserved along each neutrino trajectory, means that the operator $d/ds$ on the left-hand side of Eqn.~\ref{eq11} can be approximated as $d/dR + O(1/R^2)$.  Since, in the small-angle approximation, we are keeping terms to leading order, we will use $d\hat{\rho}/ds \approx d\hat{\rho}/dR \equiv \dot{\hat{\rho}}$.

\subsection{Calculation of the matter potential}
The components of the matter potential are given by Eq.~\ref{eq:Hm}.  To explicitly calculate the matter potential from this expression, we must determine the electron fraction, which in turn depends on the neutrino and anti-neutrino flux and flavor content, as well as on the history of the system's evolution. A realistic calculation of matter composition is a complicated problem, and a sophisticated treatment is beyond the scope of our paper.  Instead, we qualitatively estimate the electron fraction by making several simplifying assumptions.

We first assume that matter consists of only neutrons, protons and leptons and that it is in equilibrium (that is, the matter is stationary, and the proton ratio at a given radius does not change over time).  These assumptions lead to the rate balance equation:
\begin{eqnarray}
\Gamma_{pe^-\rightarrow n\nu}+\Gamma_{p\bar{\nu}\rightarrow ne^+}=\Gamma_{n\nu\rightarrow pe^-}+\Gamma_{ne^+\rightarrow p\bar{\nu}}
\end{eqnarray}

Together with neutrino spectra, baryon number density, entropy per baryon, and the constraint of charge neutrality, the rate balance equation can be solved for the equilibrium value of $Y_e$.  

Here, we are using a single-energy model of neutrino spectra, so we do not have the full energy-dependent neutrino distributions.  To approximately reconstruct the spectra that enter into the energy balance equation, we use our assumption that neutrinos and anti-neutrinos at all energies transform in the same way.  Then, flavor evolution simply leads to a partial swap of initial electron and $x-$flavor spectra, to an extent given by the neutrino and anti-neutrino survival probabilities $P_S$ and $\bar{P}_S$:
\begin{eqnarray}
f_e(E,R) &=& P_S(R)f_e(E,R_0)+(1-P_S(R))f_x(E,R_0)
\nonumber\\ 
\bar{f}_e(E,R) &=& \bar{P}_S(R)\bar{f}_e(E,R_0)+(1-\bar{P}_S(R))\bar{f}_x(E,R_0)\ \ \ \ \ \ 
\label{eq26}
\end{eqnarray}

For the initial spectra at $R = R_0$, we use Fermi-Dirac-like distributions given in terms of parameters $X$ and $T$:
\begin{eqnarray}
f_i\left(E\right)=\left(1+e^{E/T-X}\right)^{-1}
\end{eqnarray}

As an example, in our benchmark model presented below, we use the following parameters for electron neutrinos, electron anti-neutrinos, and neutrinos and anti-neutrinos of other flavors:
\begin{eqnarray}
T_e = 3.5\,{\rm MeV}\ \ \ \ \ \ X_e = 2.5\nonumber\\
T_{\bar{e}} = 4.0\,{\rm MeV}\ \ \ \ \ \ X_{\bar{e}}=2.3\nonumber\\
T_x = 4.5\,{\rm MeV}\ \ \ \ \ \ X_{x}=0.0
\label{eq22}
\end{eqnarray}
This gives a value of $\alpha = 1.4$ and a neutrino potential of $4.6\times 10^5$ ${\rm km}^{-1}$ at the neutrinosphere.

Since we are only seeking a qualitative approximation of $Y_e$, we can make further simplifications.  At the neutrino fluxes and matter densities used in our model with $R_{\rm MNR} = 60$ km, the rate balance equation in the vicinity of the MNR transition is dominated by neutrino and anti-neutrino capture, provided that the entropy per baryon ratio is sufficiently high ($\sim 50k_B$).  Electron capture produces a small shift in $Y_e$ towards lower values at smaller radius, but this effect has little impact on the qualitative features of the solution, so we neglect electron and positron capture processes and retain only the neutrino feedback.  The rate balance equation then becomes
\begin{eqnarray}
\Gamma_{p\bar{\nu}\rightarrow ne^+}\approx\Gamma_{n\nu\rightarrow pe^-}
\end{eqnarray}

The capture rate on protons is proportional to the proton number density, $n_p = n_B Y_e$, and the capture rate on neutrons is proportional to the neutron number density, $n_n = n_B (1-Y_e)$.  Both sides of the equation are also proportional to the total neutrino density, which decreases with radius as $R_{\rm NS}^2/R^2$.  Therefore, we can factor out $n_B R_{\rm NS}^2/R^2$ and write
\begin{eqnarray}
Y_e\gamma_{p\bar{\nu}\rightarrow ne^+}=\left(1-Y_e\right)\gamma_{n\nu\rightarrow pe^-}
\label{eq29}
\end{eqnarray}

where the quantities $\gamma$ are simply proportional to total scattering cross-sections for neutrinos on protons or neutrons, multiplied by the neutrino spectra, integrated over energy and angle:
\begin{eqnarray}
\gamma_{n\nu\rightarrow pe^-}\propto\int E^2 dE f_e\left(E,R\right) \sigma_{n\nu\rightarrow pe^-}\left(E\right)
\nonumber\\
\gamma_{p\bar{\nu}\rightarrow ne^+}\propto\int E^2 dE \bar{f}_e\left(E,R\right) \sigma_{p\bar{\nu}\rightarrow ne^+}\left(E\right)
\end{eqnarray}

Substituting the expression for the spectra from Eq.~\ref{eq26} then gives
\begin{eqnarray}
\gamma_{n\nu\rightarrow pe^-}= P_S\gamma_e+\left(1-P_S\right)\gamma_x
\nonumber\\
\gamma_{p\bar{\nu}\rightarrow ne^+}= \bar{P}_S\bar{\gamma}_e+\left(1-\bar{P}_S\right)\bar{\gamma}_x
\label{eq32}
\end{eqnarray}
where
\begin{eqnarray}
\gamma_{e(x)}&\propto&\int_0^\infty E^2 dE f_{e(x)}\left(E,R_0\right) \sigma_{n\nu\rightarrow pe^-}
\nonumber\\
\bar{\gamma}_{e(x)}&\propto&\int_{Q+m_e}^\infty E^2 dE \bar{f}_{e(x)}\left(E,R_0\right) \sigma_{p\bar{\nu}\rightarrow ne^+}
\end{eqnarray}
where $Q$ is the neutron - proton mass difference.  Since we only need the ratios of the $\gamma$ quantities, we can drop the common factors of coupling constants from the cross-sections and only retain the energy dependence:
\begin{eqnarray}
 \sigma_{n\nu\rightarrow pe^-}&\propto& E^2\left(1+\frac{Q}{E}\right)\sqrt{1+2\frac{Q}{E}+\frac{Q^2-m_e^2}{E^2}}
\nonumber\\
\sigma_{p\bar{\nu}\rightarrow ne^+}&\propto& E^2\left(1-\frac{Q}{E}\right)\sqrt{1-2\frac{Q}{E}+\frac{Q^2-m_e^2}{E^2}}
\end{eqnarray}

Note that $\gamma_e$, $\gamma_x$, $\bar{\gamma}_e$ and $\bar{\gamma}_x$ are all constant with radius, so the only radial dependence of $Y_e$ comes from changes in the neutrino and anti-neutrino survival probabilities.  Solving Eqn.~\ref{eq29} for $Y_e$ and substituting Eqn.\ref{eq32} gives
\begin{eqnarray}
Y_e = \frac{P_S\gamma_e+\left(1-P_S\right)\gamma_x}{P_S\gamma_e+\left(1-P_S\right)\gamma_x+\bar{P}_S\bar{\gamma}_e+\left(1-\bar{P}_S\right)\bar{\gamma}_x}
\end{eqnarray}

Since the $\gamma$ quantities appear in every term in the numerator and denominator, they are defined only up to a common proportionality constant.  Choosing the constant so that $\gamma_e-\gamma_x = 1$, we obtain $\gamma_e = 1.51$, $\gamma_x = 0.51$, $\bar{\gamma}_e = 1.83$, $\bar{\gamma}_x = 0.39$ for the choice of parameters given by Eq.~\ref{eq22}.  Then, we obtain the following expression for $Y_e$:
\begin{eqnarray}
Y_e = \frac{P_S+0.51}{P_S+1.44\bar{P}_S+0.90}
\label{eq:yeps}
\end{eqnarray}
This gives a value of $Y_e = 0.45$ for untransformed neutrino spectra, which is on the higher end of $Y_e\approx 0.35-0.45$ seen in this region in merger simulations.  The slight over-estimate of $Y_e$ is largely due to neglecting electron absorption, which, albeit subdominant at these densities and neutrino fluxes, would drive $Y_e$ a few percent lower.

The final parameter in the matter potential is the baryon number density at the neutrinosphere, $n_{\rm B,NS}$.  Here, we set this value to give a desired MNR radius, $R_{\rm MNR}$, defined as the radius at which the radially-emitted neutrinos first cross the resonance.  For our benchmark model, we choose a value of $n_{\rm B,NS}$ that gives $R_{\rm MNR} = 60$ km, which corresponds to $H_M\left(R_{\rm NS}\right) = 1.15\times10^5\ {\rm km}^{-1}$ or $n_{\rm B, NS} = 3.9\times 10^{32}\ {\rm cm}^{-3}$.  By the onset of the MNR transition, the value of $n_B$ decreases to $6.1\times 10^{30}\ {\rm cm}^{-3}$.

\subsection{Numerical implementation}

To numerically solve the integro-differential equations \ref{eq11}, we bin neutrinos by equal intervals of $v=u_0^2$ and arrive at a set of ODEs:
\begin{eqnarray}
\dot{\hat{\rho}}_i &=& \left[\vec{H}_M-\vec{H}_V+k\frac{R_{\rm NS}^4}{2R^4}\left[\left(2-v_i\right)\vec{\Phi}_1-\vec{\Phi}_3\right]\right]\times\hat{\rho}_i
\nonumber\\
\dot{\hat{\bar{\rho}}}_i &=& \left[\vec{H}_M+\vec{\bar{H}}_V+k\frac{R_{\rm NS}^4}{2R^4}\left[\left(2-v_i\right)\vec{\Phi}_1-\vec{\Phi}_3\right]\right]\times\hat{\bar{\rho}}_i
\label{eq19}
\end{eqnarray}
where the dot indicates a derivative with respect to $R$, the subscript $i$ is an integer label indicating a particular angular bin, and the moments are calculated as follows:
\begin{eqnarray}
\Phi_1 &=& 
\frac{1}{2N}\sum_j\left[\hat{\rho}_j-\alpha\bar{\hat{\rho}}_j\right]
\nonumber\\
\Phi_3 &=& 
\frac{1}{2N}\sum_j v_j\left[\hat{\rho}_j-\alpha\bar{\hat{\rho}}_j\right]
\end{eqnarray}
These moments are the same for all angular bins, and therefore must be calculated only once per step.  While, for three or more flavors of neutrinos, it can be important to use integrators that explicitly preserve unitarity, the two-flavor system of equations (\ref{eq19}) can usually be integrated via common adaptive Runge-Kutta methods.  However, in some situations, particularly with MNR where there is a near-cancelation between the neutrino and matter terms, the equations become stiff and the adaptive step size becomes prohibitively small.  At points where stiffness occurs, it can be advantageous to switch to implicit integrators, which, although far more computationally expensive per step, are able to take much larger steps.

The number of angular bins necessary for convergence varies with conditions.  We check for convergence by repeating each run at a lower angular resolution, and, if the results of the calculation (the evolution of the Hamiltonian and the survival probabilities) remain the same, we conclude that the angular resolution is sufficiently high.  We find that typically a rather high number of angular bins, of $O(4000)$ or more, is required, especially under conditions where MNR occurs at a high matter and neutrino density.  This is largely because an accurate calculation of the MNR crossing requires multiple angular bins to be near resonance at the same time, to avoid stochastic noise from individual bins going on and off resonance.  At high densities, the resonance is very narrow, so only a small fraction of angular bins are near resonance at any given time, and the overall number of angular bins must be very large.  This difficulty could perhaps be addressed in the future via dynamical binning schemes ({\it e.g.,} adaptive mesh refinement), but at present we continue to use static binning with a large number of angular bins.

\section{Results}
\label{sec:results}
\subsection{Comparison of multi-angle to  single-angle model}
We use the model described in the previous section wth parameter choices that are guided by the physical scales in merger scenarios.   For our benchmark example, the specific parameters we choose are $\alpha = 1.4$ and $R_{\rm NS} = 15 \ {\rm km}$, with $Y_e$ response to neutrino flavor transformation given by Eq.~\ref{eq:yeps}.  The radially-emitted neutrinos cross the matter-neutrino resonance at $R_{MNR} = 60\ {\rm km}$, which is roughly the scale at which the MNR transformation takes place in the dynamical calculation of Ref. \cite{Zhu:2016mwa}.  We compare the results of our single-energy, multi-angle bulb calculation with the results of a single-angle calculation, in which all neutrinos cross the resonance at $R_{MNR} = 60$ km. 

\begin{figure}
\includegraphics[width=2.8in]{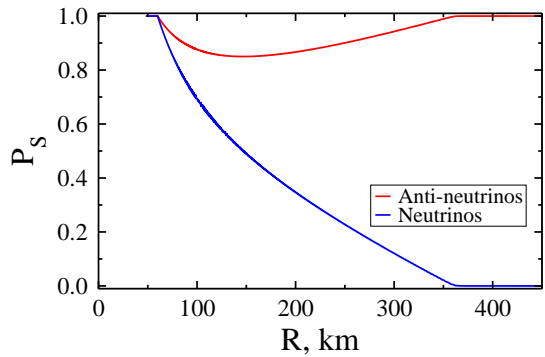}
\caption{Survival probabilities for neutrinos and anti-neutrinos in the single-energy, single-angle calculation, as a function of radius.}
\label{psingle}
\end{figure}

\begin{figure}
\includegraphics[width=2.8in]{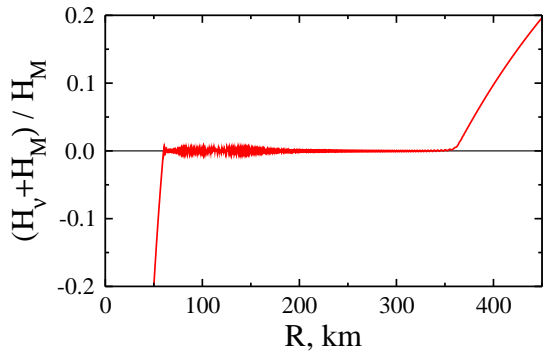}
\caption{Ratio of flavor-diagonal part of total potential $H=H_{\nu}+H_M$ to matter potential $H_M$ in the single-energy, single-angle calculation.}
\label{hsingle}
\end{figure}

Fig.~\ref{psingle} shows the survival probabilities for neutrinos, $P_{\nu_e} = \hat{\rho}_{3} + 1/2$ and anti-neutrinos $P_{\bar{\nu}_e} = \hat{\bar{\rho}}_3 + 1/2$ in the single-angle calculation.   As seen in previous single-angle MNR calculations, at first, both neutrinos and anti-neutrinos transform, but then the anti-neutrinos return to the initial flavor state while the neutrinos continue to transform until the flavor is completely swapped.  This results in a feedback effect which keeps the matter plus neutrino potential close to zero, so that the system remains near resonance until the flavor transformation is complete.  Fig.~\ref{hsingle} shows the evolution of the flavor-diagonal part of the total potential, $H_{\nu \nu, 3} + H_{M,3}$  relative to the matter potential, $H_{M,3}$ for the single angle case.

\begin{figure}
\includegraphics[width=2.8in]{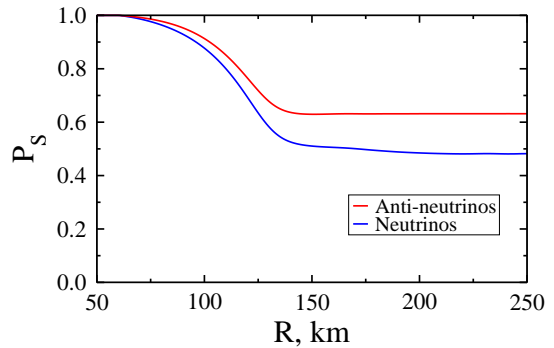}
\caption{Angle-averaged survival probabilities for neutrinos and anti-neutrinos in the multi-angle, single-energy calculation with $R_{MNR}=100$ km, as a function of radius.}
\label{spt14-15-60}
\end{figure}

Fig.~\ref{spt14-15-60} shows the angle averaged survival probabilities for the multi-angle calculation.  Unlike in the single-angle calculation, the anti-neutrino survival probability does not return to 1, and the neutrinos do not completely transform.  However, like in the single-angle case, neutrinos transform to a greater extent than anti-neutrinos. Comparing with Fig. \ref{psingle} we see that the range of radius over which flavor transformation occurs is smaller in the multi-angle case than it is in the single-angle case, $\sim 100\ {\rm km}$ instead of $\sim 300\ {\rm km}$. 

\begin{figure}
\includegraphics[width=2.8in]{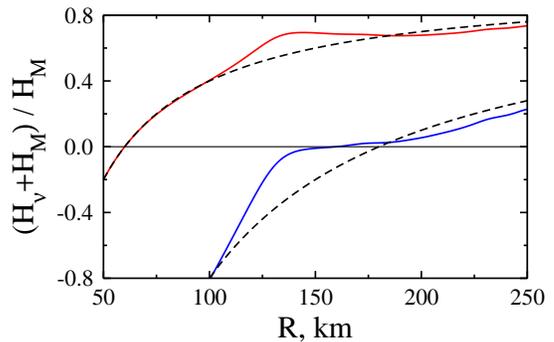}
\caption{Ratio of flavor-diagonal part of total potential $H=H_{\nu, 3}+H_M$ to matter potential $H_M$ for the radially-emitted bin (upper curve) and the tangentially-emitted bin (lower curve).   Curves for intermediate bins lie between these two extremes.  Dashed lines show the ratio without flavor transformation.}
\label{h14-15-60}
\end{figure}

Fig.~\ref{h14-15-60} shows the evolution of the flavor-diagonal component of the matter plus neutrino potentialpotential for the radially-emitted bin ($u_0= 0$) and the tangentially-emitted bin, ($u_0 = 1$) in the multi-angle calculation.  There is a range in radius, between about $120$ km and $200$ km, at which the multi-angle potential remains relatively flat, similar to what is seen in the single-angle case.  However, compared to the single-angle case, this `plateau' is of limited extent, and comparing with Fig. \ref{spt14-15-60}  we see that it occurs toward the end of the transformation.  Much of the transformation occurs earlier  at $60$ km $< R < 120$ km and interestingly, the net effect on the potential is opposite to what is seen in the single angle MNR.  Instead of a ``plateau'', the potential changes more steeply than it would if no transformation were to take place.  Because of this rapid change, the  angular bins which cross the resonance in that region spend relatively little time near resonance.  Those that cross in the ``plateau'' region will spend more time near the resonance, allowing for a feedback loop similar to the single angle case.

\subsection{Properties of the multi-angle MNR solution}
\begin{figure}
\includegraphics[width=2.8in]{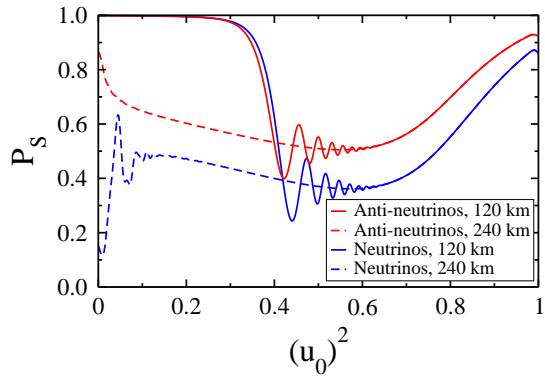}
\caption{Survival probabilities as a function of emission angle squared at $R = 120$ km and $240$ km.}
\label{spu14-15-60}
\end{figure}

As noted in Sec. \ref{sec:description}, a key feature of the multi-angle bulb model is that different angular bins pass through resonance at different times.  Fig.~\ref{spu14-15-60} shows the survival probabilities as a function of angle at $R = 120$ km, about halfway through the transformation, as well as at
$R = 240$ km, after all bins have passed through the resonance.  At $R = 120$ km, the more tangentially emitted bins, with $u^2 < 0.4$, have not yet passed through resonance, so they are untransformed.  Bins with $0.4 < u^2 < 0.6$ have recently gone through resonance and are undergoing flavor transformation, while bins with $u^2 > 0.6$ have moved far away from resonance, completed their flavor transformation and approached the final flavor state seen in the $R = 240$ km curve. Looking at this final flavor state curve, we see that survival probabilities vary with angle, but in general, neutrinos transform more than anti-neutrinos in all angular bins.

In the single-angle model, the magnitude of the neutrino potential is always enhanced by the flavor transformation.  Because the untransformed neutrino potential drops off faster than the matter potential, this enhancement allows the system to maintain a cancelation between the neutrino and matter potentials and remain on resonance for a wide range of $R$.  However, in the multi-angle model, the magnitude of the neutrino potential is at first suppressed, causing the total potential to rise and the angular bins to pass through resonance faster.  This is seen in Fig.~\ref{h14-15-60} for the range $100$ km$<R<120$ km.    However, at $R > 120$ km, when the most tangential bins begin to reach the resonance, the neutrino potential is again enhanced, just like in the single-flavor case.

\begin{figure}
\includegraphics[width=2.8in]{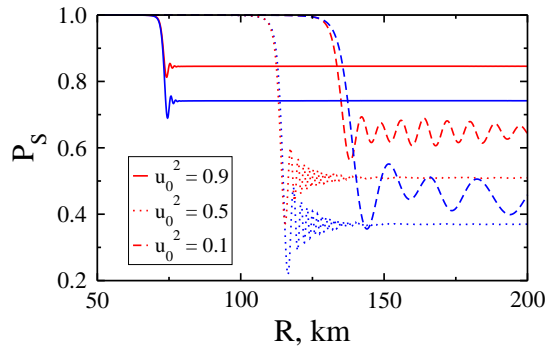}
\caption{Survival probabilities for neutrinos and anti-neutrinos on trajectories with $u_0^2 = \cos^2\theta_0 = 0.9, 0.5$ and $0.1$ as a function of radius.}
\label{spbin14-15-60}
\end{figure}

This phenomenon may be related to the pattern of flavor transformation of each individual angular bin. Fig.~\ref{spbin14-15-60} shows the flavor evolution of several angular bins.  We see that initially, when the bin has just passed through resonance, both neutrinos and anti-neutrinos transform in the same way.  This is to be expected, because most of this transformation takes place somewhat after the bin has passed through resonance, so that the matter plus neutrino potential is already much larger than the vacuum term.  Because the only difference between the neutrino and anti-neutrino Hamiltonian is the vacuum term, neutrinos and anti-neutrinos initially have almost the same flavor evolution.  However, after some time, enough difference accumulates between the neutrino and anti-neutrino flavor vectors that the two begin to separate, just like in single-angle MNR.  But unlike in single-angle MNR, the bin in question is moving rapidly away from resonance, so the transformation does not have time to complete.  Thus neutrinos do not transform completely, and the anti-neutrinos do not return to the initial flavor state.

When neutrinos and anti-neutrinos transform identically, the magnitude of the neutrino potential is suppressed by the flavor transformation, since the contribution to the potential from this bin is proportional to $\hat{\rho}-\alpha\hat{\bar{\rho}}$.  However, when only the neutrinos transform, the electron flavor component of the neutrino potential becomes more negative.  Thus, the flavor transformation in each bin contributes to suppression of the neutrino potential in the early stages, but enhances the neutrino potential later on.  

To obtain the overall effect on the neutrino potential, we can average over all angular bins that are undergoing flavor transformation at a given radius.  Early on, the suppression effect predominates.  However, near the end of the flavor transformation, when there are only a few bins remaining near resonance and they are all in the later stage of their flavor evolution, the enhancement effect is greater.  In this region, the only neutrinos remaining near resonance are the almost tangentially-emitted ones, all propagating at nearly the same trajectory angle, and the situation begins to resemble single-angle MNR.

\begin{figure}
\includegraphics[width=2.8in]{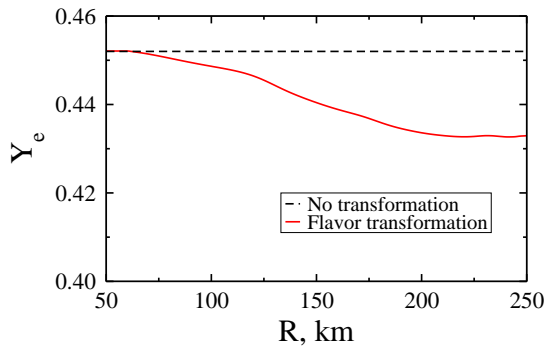}
\caption{Effect of flavor transformation on electron fraction as a function of radius.}
\label{ye14-15-60}
\end{figure}

Finally, Fig.~\ref{ye14-15-60} shows the effect of neutrino flavor transformation on the electron fraction $Y_e$.  In the approximation made here, that matter is in equilibrium with neutrinos and electron capture reaction rates are negligible, $Y_e$ depends only on neutrino and anti-neutrino survival probabilities, and remains constant in the absense of flavor transformation.  We see that flavor transformation decreases $Y_e$ by a small amount ($\sim 5$ percent).  While this effect is modest compared to what one would obtain with single-angle MNR, even small changes in $Y_e$ can have a strong effect on nucleosynthesis.

\subsection{Dependence of solution on matter and neutrino densities}
The calculation described above was performed using the parameters $R_{\rm NS} = 15$ km, $\alpha = 1.4$ and $R_{\rm MNR} = 60$ km.  To determine whether the phenomena we found are robust under different conditions, we perform several calculations with varying parameters.  

First, we perform two calculations with different choices of neutrinosphere radius, $R_{\rm NS} = 10$ km and $20$ km.  A larger neutrinosphere radius corresponds to a higher flux (by a factor of $R_{\rm NS}^2$) and  larger opening angle between neutrino trajectories (also by a factor of $R_{\rm NS}^2$ at a given radius), and therefore a larger neutrino potential relative to the vacuum term.  The matter potential is adjusted to keep $R_{\rm MNR}$ fixed at 60 km.

\begin{figure}
\includegraphics[width=2.8in]{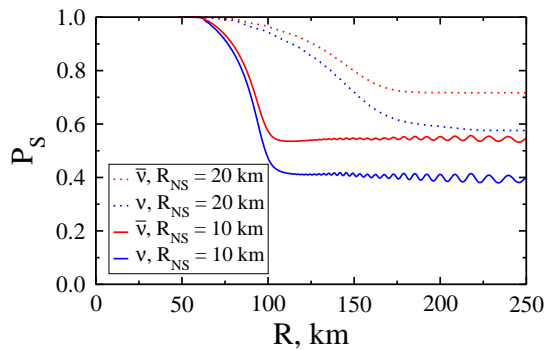}
\caption{Survival probabilities for different choices of neutrinosphere radius, $R_{\rm NS} = 10$ km and $20$ km, with fixed $\alpha = 1.4$ and $R_{\rm MNR}=60$ km.}
\label{sp14-1020-60}
\end{figure}

The evolution of total survival probabilities from these two calculations are shown in Fig.~\ref{sp14-1020-60}.  As expected, a larger neutrinosphere radius partially suppresses flavor transformation:  there is less flavor transformation overall, and it takes longer to complete.  This can be attributed to two effects:  first, a larger $R_{\rm NS}$ leads to a larger opening angle between neutrino trajectories at a given radius.  Therefore, a smaller fraction of angular bins is on resonance at any given time.  Second, for a larger $R_{\rm NS}$, the ratio of the neutrino and matter potentials to the vacuum mixing term is larger.  Therefore, the in-medium mixing angle is smaller, and flavor transformation is suppressed.

Next, we examine the effects of changing the MNR radius.  As discussed above, this parameter is related to the density of the matter profile:  for a given neutrino flux, a smaller $R_{\rm MNR}$ corresponds to a higher density of matter.  We perform two calculations with $R_{\rm MNR} = 40$ km and $80$ km, both with $\alpha = 1.4$ and $R_{\rm NS} = 15$ km.

\begin{figure}
\includegraphics[width=2.8in]{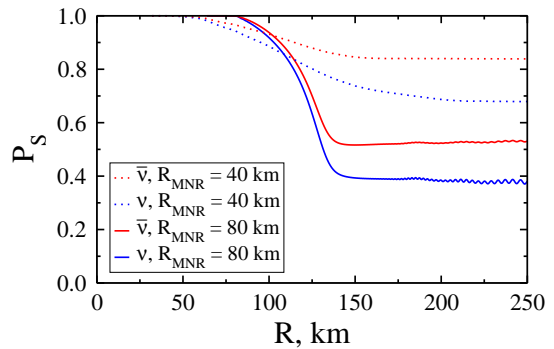}
\caption{Survival probabilities for different choices of the MNR radius, $R_{\rm MNR} = 40$ and $80$ km, with fixed $\alpha = 1.4$ and $R_{\rm NS}=15$ km.}
\label{sp14-15-4080}
\end{figure}

Survival probabilities from these calculations are shown in Fig.~\ref{sp14-15-4080}.  We see that flavor transformation is less efficient for a smaller choice MNR radius.  The mechanism is similar to the suppression of flavor evolution by a large neutrinosphere radius:  because the MNR transition begins at a smaller radius, the neutrino and matter potential during the transition is larger relative to the vacuum mixing term, and the opening angle between neutrino trajectories is also larger.  In addition, the region in which MNR can take place is smaller for a smaller choice of MNR radius, so there is less time available for flavor transformation.

\begin{figure}
\includegraphics[width=2.8in]{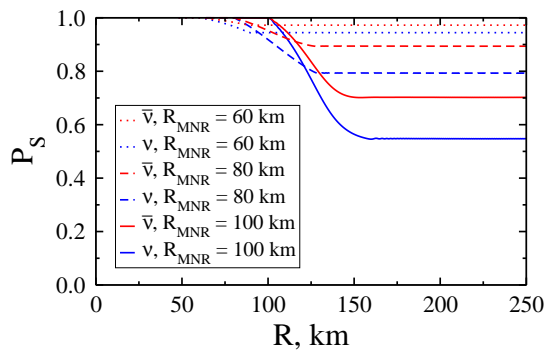}
\caption{Survival probabilities with matter density profile $M \propto R^{-1}$, for different choices of the MNR radius, $R_{\rm MNR} = 40, 80$ and $100$ km, with fixed $\alpha = 1.4$ and $R_{\rm NS}=15$ km.}
\label{sp14-15-M1}
\end{figure}

We next change the rate at which the matter density drops off with radius.  Fig.~\ref{sp14-15-M1} shows the evolution of survival probabilities with a slowly-decreasing matter potential, proportional to $R^{-1}$ instead of $R^{-3}$.  We see that a slowly-decreasing matter potential strongly suppresses flavor transformation compared to the rapidly-decreasing case.  For $R_{\rm MNR} = 60$ km, $R_{\rm NS} = 15$ km and $\alpha = 1.4$, there is almost no flavor transformation.  However, increasing the MNR radius to 80-100 km restores the pattern of flavor transformation seen in our other calculations. 

Next, we examine the effects of changing the parameter $\alpha$, the initial ratio of the anti-neutrino to the neutrino contribution to the potential.  We perform two calculations with $\alpha = 1.2$ and $1.6$, both with $R_{\rm NS} = 15$ km and $R_{\rm MNR} = 60$ km. 

\begin{figure}
\includegraphics[width=2.8in]{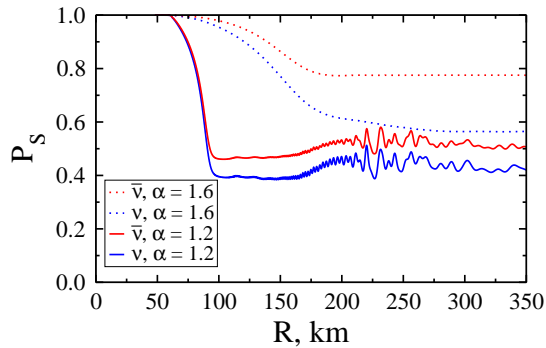}
\caption{Survival probabilities for different choices of initial neutrino spectra, with $\alpha = 1.2$ and $1.6$, at fixed $R_{\rm NS}=15$ km and $R_{\rm MNR}=60$ km.}
\label{sp1216-15-60}
\end{figure}

The results of the calculations with varying values of $\alpha$ are shown in Fig.~\ref{sp1216-15-60}.  While the smaller value of $\alpha$ exhibits greater and more rapid flavor conversion, the larger value of $\alpha$ gives a greater separation between flavor evolution of neutrinos and anti-neutrinos.  This is because, for a larger value of $\alpha$, the neutrino potential is higher at a given radius, suppressing flavor evolution to some extent.  However, due to the greater difference between neutrino and anti-neutrino spectra, the system at high $\alpha$ has a greater capacity to sustain the MNR, and so there is more time for the separation between neutrino and anti-neutrino flavor to develop.  Note that the horizontal scale of Fig.~\ref{sp1216-15-60} is larger than that of the previous two figures; this is because the MNR is sustained for a larger range of radius in the $\alpha = 1.6$ case.

\begin{figure}
\includegraphics[width=2.8in]{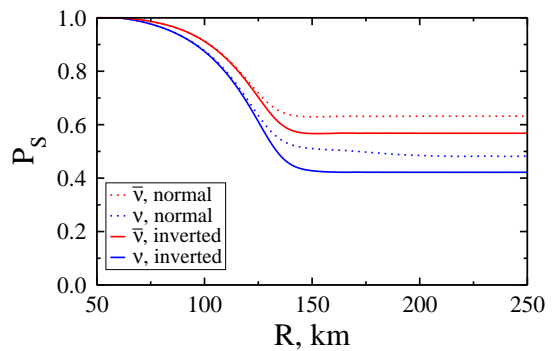}
\caption{Survival probabilities for different choices of mass hierarchy, with $\alpha = 1.4$, $R_{\rm NS}=15$ km and $R_{\rm MNR}=60$ km.}
\label{sp14-15-60inv}
\end{figure}

Finally, we examine the effects of changing the mass hierarchy.  Fig.~\ref{sp14-15-60inv} shows the evolution of survival probabilities for the normal and the inverted hierarchy, with $\alpha = 1.4$, , $R_{\rm NS}=15$ km and $R_{\rm MNR}=60$ km.  We see that in the inverted hierarchy, there is slightly more flavor transformation, but otherwise, the flavor evolution is very similar.

The pattern of flavor evolution in all these simulations is qualitatively similar.  Both neutrinos and anti-neutrinos transform to some extent, with neutrinos undergoing more flavor transformation than anti-neutrinos.

\subsection{Effects of additional neutrino emission from $R > R_{\rm NS}$}
Thus far, we considered a situation where all neutrinos are emitted at $R = R_{\rm NS}$ with a semi-isotropic angular distribution.  However, in a real compact object merger, the neutrinosphere may not be sharply defined, and in addition a significant portion of neutrinos is emitted from an accretion disk at $R > R_{\rm NS}$. There may also be a scattered halo of neutrinos that have undergone non-forward scattering by dense matter high above the neutrinosphere.  While, for $R > R_{\rm NS}$, the neutrinos emitted at $R_{\rm NS}$ are confined to a narrow range of trajectory angles, the trajectories of these additional neutrinos cover a much wider angular range.  For this reason, neutrinos emitted or scattered at larger $R$ contribute disproportionately to the neutrino potential.  We would like to examine how the presence of these neutrinos affects our results.

Current state of the art multi-angle simulations of neutrino flavor transformation are performed in spherical symmetry, but we are nevertheless able to incorporate additional sources of neutrinos in our model to assess the importance of a more extended source of neutrinos. To do this, we introduce the idea of an extended neutrinosphere 
\cite{Cherry:2013mv}. In short, we keep the original spherical emitted surface, but above it add to it an emitting volume of neutrinos.  The amount of emission from this extended volume decreases with increasing distance from the center of the object.  We describe this model in more detail in the next few paragraphs. 

Suppose that instead of specifying the initial conditions at $R_{\rm NS}$, we specify them at an extended radius $R_{\rm E}> R_{\rm NS}$.  In doing so, we assume that there is no flavor transformation between $R_{\rm NS}$ and $R_{\rm E}$.  Instead of labeling neutrinos by the cosine of the emission angle at the neutrinosphere, $u_0$, we label them by the cosine of the trajectory angle at $R = R_{\rm E}$, $u_E \equiv u\left(R_E\right)$.  We also assume that there is no significant neutrino emission or non-forward scattering above $R_E$.

First, consider the semi-isotropic bulb model that we have been working with thus far.  Here, the timelike and radial components of the neutrino potential at $R = R_E$ are given by Eq.~\ref{eq:H0HR}.  Adopting the single-energy approximation and dropping proporitonality constants, we have
\begin{eqnarray}
\vec{H}_0\left(R_E\right) &\propto& \int_{u_E^{\rm min}}^1 du_E^{\prime}\left(\hat{\rho}-\alpha\hat{\bar{\rho}}\right)
\nonumber\\
\vec{H}_R\left(R_E\right) &\propto& \int_{u_E^{\rm min}}^1 u_E^{\prime}du_E^{\prime}\left(\hat{\rho}-\alpha\hat{\bar{\rho}}\right)
\end{eqnarray}
where $u_E^{\rm min} = \sqrt{1-R_{\rm NS}^2/R_E^2}$.  The lower limit of integration can be set to 0 by using the Heaviside step function:
\begin{eqnarray}
\vec{H}_0\left(R_E\right) &\propto& \int_0^1 du_E^{\prime}\theta\left(u_E-u_E^{\rm min}\right)\left(\hat{\rho}-\alpha\hat{\bar{\rho}}\right)
\nonumber\\
\vec{H}_R\left(R_E\right) &\propto& \int_0^1 u_E^{\prime}du_E^{\prime}\theta\left(u_E-u_E^{\rm min}\right)\left(\hat{\rho}-\alpha\hat{\bar{\rho}}\right)
\end{eqnarray}
Following the analysis of Sec. II.A., Eq.~\ref{eq:H0HR}-\ref{eq9}, these quantities at any $R > R_E$ can be written as integrals over $u_E$ instead of $u\left(R\right)$:
\begin{eqnarray}
\vec{H}_0\left(R\right)&\propto&\int_0^1\frac{u_E^{\prime}}{u^\prime}du_E^{\prime}\theta\left(u_E-u_E^{\rm min}\right)\left(\hat{\rho}-\alpha\hat{\bar{\rho}}\right)
\nonumber\\
\vec{H}_R\left(R\right)&\propto&\int_0^1 u_E^{\prime}du_E^{\prime}\theta\left(u_E-u_E^{\rm min}\right)\left(\hat{\rho}-\alpha\hat{\bar{\rho}}\right)
\label{eq:extbulb}
\end{eqnarray}
where
\begin{eqnarray}
u = \sqrt{1-\frac{R_E^2}{R^2}\left(1-u_E^2\right)}
\end{eqnarray}
So far, this is a trivial re-parametrizaton of the original bulb problem.  In other words, as long as there is no flavor transformation between $R_{\rm NS}$ and $R_{\rm E}$, we can change the emission radius from $R_{\rm NS}$ to $R_{\rm E}$, replace the quantities describing neutrino flavor $\hat{\rho}-\alpha\hat{\bar{\rho}}$ with $\theta\left(u_E-u_E^{\rm min}\right)\left(\hat{\rho}-\alpha\hat{\bar{\rho}}\right)$ (so that no neutrinos have trajectories shallower than $u_E^{\rm min}$ at $R = R_{\rm E}$) and relabel neutrino trajectories by $u_E \equiv u\left(R_E\right)$ instead of $u_0\equiv u\left(R_{\rm NS}\right)$, leaving the model unchanged.

Having reformulated the problem in this way, we can now add additional neutrinos emitted above $R_{\rm NS}$ but below $R_E$.  To do this, we 
replace the Heaviside step function with a general weight function $F\left(u_E\right)$:
\begin{eqnarray}
\theta\left(u_E-u_E^{\rm min}\right)\left(\hat{\rho}-\alpha\hat{\bar{\rho}}\right)\rightarrow F\left(u_E\right)\left(\hat{\rho}-\alpha\hat{\bar{\rho}}\right)
\end{eqnarray}

There are now neutrinos traveling on all trajectories for which $F$ is nonzero, with the value of $F$ determining the relative number of neutrinos on a particular trajectory.  In principle, we could also make the parameter $\alpha$ also depend on $u_E$, or, equivalently, use different weight functions for $\hat{\rho}$ and $\hat{\bar{\rho}}$.  This would correspond to a situation where the emission profile for neutrinos is different from that for anti-neutrinos.  While such a situation is realistic, for the sake of clarity, we do not consider it here, and make the assumption that neutrinos and anti-neutrinos are emitted in equal proportion on all trajectories.

Next, we choose the function $F$.  Our goal is simply to examine the effect of having additional neutrinos traveling on trajectories with larger angles, so any convenient function that provides these while remaining physically and mathematically reasonable will suffice.  We keep $F\left(1\right) = 1$ and $F'\left(1\right) = 0$, as in the neutrino bulb model.  We assume that neutrino emission goes to zero as $R\rightarrow R_E$, which gives $F\left(0\right) = 0$.  We would like $F$ to be a monotonically increasing function of $u_E$, and to have at least one free parameter that can be adjusted so that we can choose a value for the neutrino flux.  A simple choice that satisfies these properties is
\begin{eqnarray}
F=\left(2u_E-u_E^2\right)\exp\left[-\frac{1}{2\sigma^2}\left(1-u_E\right)^2\right]
\label{eq:wfansatz}
\end{eqnarray}
where the parameter $\sigma$ can be adjusted to give a desired neutrino flux.

\begin{figure}
\includegraphics[width=2.8in]{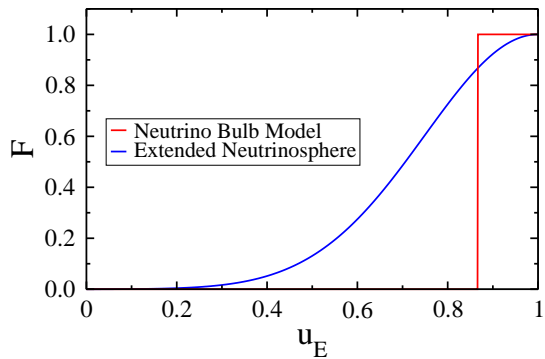}
\caption{Relative numbers of neutrinos on trajectories with $u_E = \cos\theta\left(40{\rm km}\right)$, for the standard neutrino bulb model with $R_{\rm NS}=20$ \, km, and the extended neutrinosphere model.}
\label{weights}
\end{figure}

Next, we choose specific model parameters.  We start from the bulb model, with $R_{\rm NS} = 20$ km and $\alpha = 1.4$, and choose the extended neutrinosphere radius to be $R_E = 40$ km.  This gives $u_E^{\rm min} =\sqrt{3}/2$.  The neutrino flux in the bulb model is proportional to the integral in $H_R$, the second line of Eq.~\ref{eq:extbulb}, but without the factor of $\left(\hat{\rho}-\hat{\bar{\rho}}\right)$.  For our model with extended neutrino emission, we would like the flux to be about twice as high, since roughly half the neutrino emission is expected to come from outside the proto-neutron star.  This gives the following expression that can be used to determine $\sigma$:
\begin{eqnarray}
2\int_{u_E^{\rm min}}^1 u_E^{\prime}du_E^{\prime}=\frac{1}{4}
\nonumber\\
=\int_0^1 u_E^{\prime}du_E^{\prime}\left(2u_E-u_E^{\prime 2}\right)\exp\left[-\frac{1}{2\sigma^2}\left(1-u_E^{\prime}\right)^2\right]
\end{eqnarray}
which gives $\sigma\approx 0.267$.  The angular distribution of neutrinos as a function of $u_E$ for this extended neutrinosphere model is shown in Fig.~\ref{weights}, with angular distribution for the bulb model at $R_E$ shown for comparison.

It is now straightforward to modify our calculation to include the extended neutrinosphere:  we simply replace $R_{\rm NS}$ with $R_E$, and weight the contributions from the different angular bins by the function $F$.  We first verfiy that  the use of the small angle approximation in our calculations is still valid in the context of this extended model. We find that by the time the neutrinos reach the MNR radius the small angle approximation neutrino potential by only a few percent relative to its unapproximated form.     

One major impact of using the extended neutrinosphere is that the neutrino potential is significantly enhanced, not only because we make the flux twice as large to accomodate the large-angle neutrinos, but also because the larger angle neutrinos contribute more to the potential.  Consider the neutrino potential for the radially-emitted bin ($u_0 = u_E = 1$).  From Eq.~\ref{eq:hnunu},

\begin{eqnarray}
\vec{H}_{\nu\nu}\left(u_E=1\right)=k\frac{R_E^4}{2R^4}\left[\vec{\Phi}_1-\vec{\Phi}_3\right]
\end{eqnarray}
where $\Phi_N$ now includes the weight function:
\begin{eqnarray}
\vec{\Phi}_N = \int_0^1 du^{\prime}_E u^{\prime N}_E F\left(u^{\prime}_E\right)\left(\hat{\rho}-\alpha\hat{\bar{\rho}}\right)
\end{eqnarray}

In the original bulb model, $F\left(u_E\right)$ is simply $\theta\left(u_E-u_E^{\rm min}\right)$, where, with our choice of parameters, $u_E^{\rm min}=\sqrt{3}/2$.  In the absence of flavor transformation, we obtain $\vec{\Phi}_1-\vec{\Phi}_3 = \left(1/64\right)\left(1-\alpha\right)\hat{e}_3$.  In the extended bulb model, using $F$ given by Eq.~\ref{eq:wfansatz}, we instead obtain $\vec{\Phi}_1-\vec{\Phi}_3\approx 0.0734\left(1-\alpha\right)\hat{e}_3$, an enhancement by almost a factor of 5.  To facilitate comparison with simulations presented above, we would like the value of the neutrino potential in the flavor transformation region to be comparable to these simulations, rather than much higher.  For this reason, we choose a larger value of the MNR radius, $R_{\rm MNR}=90 \, {\rm km}$, instead of $60 \, {\rm km}$ as in the benchmark simulation.  Since the neutrino potential scales as $R^{-4}$, this parameter choice sets the neutrino potential in the flavor transformation region to a value comparable with our other simulations.

\begin{figure}
\includegraphics[width=2.8in]{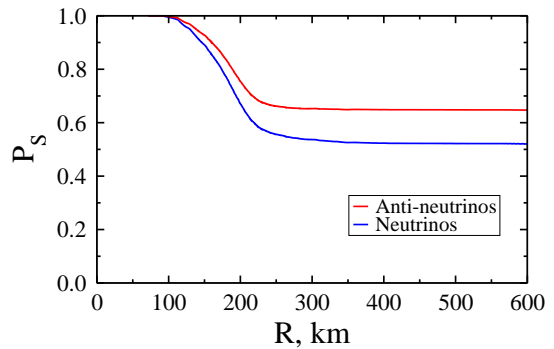}
\caption{Angle-integrated survival probabilities for neutrinos and anti-neutrinos as a function of radius in the extended neutrinosphere model.}
\label{sp14-20-90ext}
\end{figure}

The evolution of total survival probabilities for neutrinos and anti-neutrinos in the extended neutrinosphere model is shown in Fig.~\ref{sp14-20-90ext}.  The pattern of flavor evolution is similar to that seen in the neutrino bulb simulations above:  both neutrinos and anti-neutrinos partially transform, but neutrinos transform to a somewhat greater extent.  The separation between neutrino and anti-neutrino survival probabilities takes somewhat longer to develop, and is slightly less prominent, because the exponential tail in the neutrino angular distribution takes a long time to completely pass through resonance, and there are relatively few of these neutrinos, resulting in a reduced impact on survival probabilities.

\begin{figure}
\includegraphics[width=2.8in]{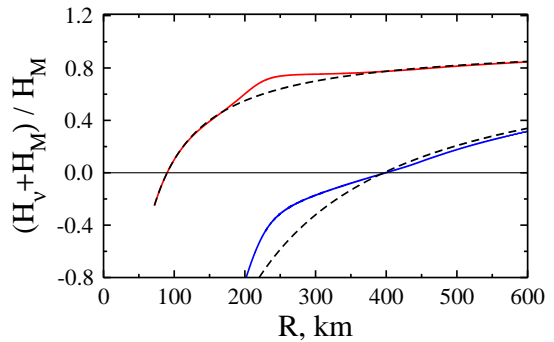}
\caption{Ratio of flavor-diagonal part of total potential to matter potential for the radially-emitted bin (upper curve) and for the most tangential bin (lower curve) in the extended neutrinosphere model.  Curves for intermediate bins lie between these two extremes.  Dashed lines show the ratio without flavor transformation.}
\label{hplot14-20-90ext}
\end{figure}

The evolution of the total potential for the radially-emitted and the most tangential bin is shown in Fig.~\ref{hplot14-20-90ext}.  Again, the qualitative behavior is similar to that seen for the bulb model in Fig.~\ref{h14-15-60}.  However, the `plateau' is not so pronounced, and is extended over a longer range ($R = 250$ km to $400$ km, instead of $125$ km to $200$ km).  This is largely due to the presence of the long tail in the neutrino angular distribution, which takes a long distance to completely pass through resonance.

\section{Conclusion}
 \label{sec:conclusions}
We have presented multi-angle calculations of neutrino flavor evolution in the spherically symmetric bulb model under MNR-like conditions.  We find that, just like in single-angle MNR calculations, the cancelation between the neutrino potential and the matter potential leads to large-scale flavor evolution in multi-angle systems.  However, while there are some shared features with the single-angle model (neutrinos transform more than anti-neutrinos and the Hamiltonian remains near zero over a certain range of radius for a small subset of angular bins) multi-angle flavor evolution is qualitatively different, so treatment of MNR-like neutrino flavor evolution in realistic systems requires multi-angle simulations.

Calculations with different model parameters indicate that increasing the neutrino flux or the matter density (increasing $R_{\rm NS}$ or decreasing $R_{\rm MNR}$) partially suppresses flavor transformation, while increasing the asymmetry between neutrino and anti-neutrino spectra (the value of $\alpha$) increases the difference in flavor evolution between neutrinos and anti-neutrinos.  In addition, adopting a very shallow matter density profile proportional to $R^{-1}$ instead of the more standard $R^{-3}$ strongly suppresses flavor transformation.  However, even in this case, a significant amount of flavor transformation still occurs provided that the MNR radius is sufficiently large ($80-100$ km).  In addition we considered a scenario with a spatially extended neutrino emission volume that declined with distance from the center of the object.  In this case, we found qualitatively similar results to the bulb model, but the transitions more extended.
 
Overall, the pattern of flavor evolution is similar under a wide range of physical conditions.  This suggests that the flavor transformation phenomena demonstrated here are robust and are likely to occur in most non-isotropic systems that allow for a cancelation between the matter and neutrino potentials.  Since this cancelation is expected to occur during certain epochs in compact object mergers, we expect this type of flavor transformation to play an important role in these environments, with effects on nucleosynthesis and potentially also other aspects of the merger.

\begin{acknowledgments} 
This material is based upon work supported by the U.S. Department of Energy, Office of Science, Office of Nuclear Physics, under Award Number DE-FG02-02ER41216.  We thank J. Kneller for useful discussions.
\end{acknowledgments}

\bibliography{compactObjectMergerSimulations,nucleosynthesisInDiskOutflow,diskOscillationCalculations,supernovaOscillationCalculations,denseMediaOscillations,misc}

\end{document}